# A relaxation function encompassing the stretched exponential and the compressed hyperbola

Mário N. Berberan-Santos

*Centro de Química-Física Molecular, Instituto Superior Técnico,*
*Universidade Técnica de Lisboa, 1049-001 Lisboa, Portugal*

*berberan@ist.utl.pt, tel. 351-218419254, fax 351-218464455*

Suggested running head: **A general relaxation function**

## Abstract

A simple relaxation function $I(t/\tau_0; \alpha, \beta)$ unifying the stretched exponential with the compressed hyperbola is obtained, and its properties studied. The scaling parameter $\tau_0$ has dimensions of time, whereas the shape-determining parameters $\alpha$ and $\beta$ are dimensionless, both taking values between 0 and 1. For short times, the relaxation function is always exponential, with time constant $\tau_0$. For small values of $\alpha$, the function is close to exponential for all times, irrespective of $\beta$. The function is also close to an exponential when $\beta$ is near unity, irrespective of $\alpha$. For large values of $\alpha$ and long times, the function is close to a stretched exponential, provided that $\beta>0$. The compressed hyperbola is recovered for $\beta=0$.

# 1. Introduction

In relaxation processes, including time-dependent luminescence spectroscopy, a perturbation is applied to the system up to an instant set as the origin of times, $t=0$. At this instant, the perturbation is suddenly removed, and the system relaxes towards the long-time (equilibrium) value. The time-dependent relaxation function $I(t)$, defined from a suitable property $P$ of the system as

$$I(t) = \frac{P(t) - P(\infty)}{P(0) - P(\infty)}, \qquad (1)$$

can also be written as

$$I(t) = \exp\left(-\int_0^t w(u)\,du\right), \qquad (2)$$

where $w(t)$ is a time-dependent rate coefficient, defined by

$$w(t) = -\frac{d \ln I}{dt}. \qquad (3)$$

In the simplest case, $w(t)$ is time-independent, and the decay is exponential. If $dw/dt > 0$, the decay is said to be super-exponential, and if $dw/dt < 0$, the decay is sub-exponential [2]. The relaxation function can also be written as:

$$I(t) = \mathcal{L}[H(k)] = \int_0^\infty H(k) e^{-kt}\,dk. \qquad (4)$$

This relation is always valid because $H(k)$ is the inverse Laplace transform of $I(t)$. The function $H(k)$ is normalized, as $I(0)=1$ implies that $\int_0^\infty H(k)\,dk = 1$. In many cases (e.g. in the absence of a rise-time), the function $H(k)$ is nonnegative for all $k>0$, and $H(k)$ can be understood as a distribution of rate constants (strictly, a probability density function) [2]. This is the situation addressed in this work.

A simple form of the inverse Laplace transform of a relaxation function can be obtained by the method outlined in [1]. Briefly, the three following equations can be used for the computation of $H(k)$ from $I(t)$,

$$H(k) = \frac{2}{\pi} \int_0^\infty \mathrm{Re}[I(i\omega)]\cos(k\omega)\,d\omega \quad k > 0, \qquad (5)$$

$$H(k) = -\frac{2}{\pi}\int_0^\infty \text{Im}[I(i\omega)]\sin(k\omega)\,d\omega \quad k > 0, \tag{6}$$

$$H(k) = \frac{1}{\pi}\int_0^\infty \left[\text{Re}[I(i\omega)]\cos(k\omega) - \text{Im}[I(i\omega)]\sin(k\omega)\right] d\omega, \tag{7}$$

it being understood that *I(z)* has no singularities for *Re(z) > 0*. Equation (7) is clearly the semi-sum of equations (5) and (6).

While there is no apparent advantage of one of the relations over the others, it turns out that for numerical integration purposes, and depending on the relaxation function and specific values of the respective parameters, some of the relations may lead to large errors whereas the other(s) yield accurate results.

**2. Stretched exponential (or Kohlrausch) function**

The stretched exponential decay function is given by

$$I(t) = \exp\left[-(t/\tau_0)^\beta\right], \tag{8}$$

where $0 < \beta \leq 1$, and $\tau_0$ is a parameter with the dimensions of time.

In studies of the relaxation of complex systems, the Kohlrausch function is frequently used as a purely empirical decay function, although there are theoretical arguments to justify its common occurrence. In the field of condensed matter luminescence, equation (8) has firm grounds on several models of luminescence quenching, namely diffusionless resonance energy transfer by the dipole-dipole mechanism, with $\beta = 1/6$, $1/3$ and $1/2$ for one-, two- and three-dimensional systems, respectively [3,4]. Other rational values of $\beta$ are obtained for different multipole interactions, e.g. $\beta = 3/8$, $3/10$, for the dipole-quadrupole and quadrupole-quadrupole mechanisms in three-dimensions [5].

The Kohlrausch decay function is convenient as a fitting function, even in the absence of a model, given that it allows gauging in simple way deviations to the "canonical" single exponential behaviour through the parameter $\beta$. This is precisely how it was introduced in relaxation studies, first by R. Kohlrausch [6] in transient electric phenomena, and then by A. Werner [7] in luminescence studies. The Kohlrausch decay function was also recently used in this way in the analysis of single-molecule fluorescence [8] and in the fluorescence lifetime imaging of biological tissues [9].

A time-dependent rate coefficient $w(t)$ can be defined for the Kohlrausch decay function by using equation (3):

$$w(t) = \frac{\beta}{\tau_0}\left(\frac{t}{\tau_0}\right)^{\beta-1}, \qquad (9)$$

where $0<\beta\leq 1$. After Williams and Watts [10], the Kohlrausch decay function is often called the "slower-than-exponential" (with respect to an exponential of lifetime $\tau_0$) function. Although sub-exponential, this is however somewhat of a misnomer, as a most characteristic aspect of the function is precisely the existence of two regimes: a faster-than-exponential (with respect to an exponential of lifetime $\tau_0$) initial decay (indeed, the rate constant is infinite for $t=0$), and a slower-than-exponential decay (with respect to an exponential of lifetime $\tau_0$) for times longer than $\tau_0$ [11]. These two regimes are very marked for small $\beta$, but become indistinct as $\beta \rightarrow 1$.

The initial, fast decaying part of the Kohlrausch function ($\beta<1$), resulting from a Lévy distribution of rate constants (see below), with its characteristic long tail, is sometimes ignored by using a $\tau_0$ smaller than the shortest time of observation, and multiplying the decay function by a factor higher than 1, a procedure that invalidates its correct normalization.

The slowing down of the decay rate can be shown explicitly by the time-dependent rate coefficient, equation (9). As mentioned, this rate coefficient is initially infinite, which is an unphysical result. In the field of energy transfer in homogeneous media, an initially infinite rate coefficient arises when point particles are assumed. If a distance of closest approach is postulated, then the initial part of the decay becomes exponential, and the decay obeys a stretched-exponential only for longer times [11]. The stretched exponential decay function is thus necessarily of an approximate nature.

The determination of $H(k)$ for a given $I(t)$ amounts to the computation of the respective inverse Laplace transform. The result, first obtained by Pollard [12], is

$$H_\beta(k) = \frac{\tau_0}{\pi}\int_0^\infty \exp(-k\tau_0 u)\exp\left[-u^\beta \cos(\beta\pi)\right]\sin\left[u^\beta \sin(\beta\pi)\right]du. \qquad (10)$$

Equivalent integral representations are obtained from equations (5)-(7) [1]:

$$H_\beta(k) = \frac{2\tau_0}{\pi} \int_0^\infty \exp\left[-u^\beta \cos\left(\frac{\beta\pi}{2}\right)\right] \cos\left[u^\beta \sin\left(\frac{\beta\pi}{2}\right)\right] \cos(k\tau_0 u)\, du \quad (k>0), \tag{11}$$

$$H_\beta(k) = \frac{2\tau_0}{\pi} \int_0^\infty \exp\left[-u^\beta \cos\left(\frac{\beta\pi}{2}\right)\right] \sin\left[u^\beta \sin\left(\frac{\beta\pi}{2}\right)\right] \sin(k\tau_0 u)\, du \quad (k>0), \tag{12}$$

$$H_\beta(k) = \frac{\tau_0}{\pi} \int_0^\infty \exp\left[-u^\beta \cos\left(\frac{\beta\pi}{2}\right)\right] \cos\left[u^\beta \sin\left(\frac{\beta\pi}{2}\right) - k\tau_0 u\right] du. \tag{13}$$

For $\beta=1$, one has of course $H_1(k)=\delta(k-1/\tau_0)$. For $\beta\neq 1$, $H_\beta(k)$ can be expressed in terms of elementary functions only for $\beta=1/2$ [12],

$$H_{1/2}(k) = \frac{\tau_0}{2\sqrt{\pi}\,(k\tau_0)^{3/2}} \exp\left(-\frac{1}{4k\tau_0}\right). \tag{14}$$

A relatively simple form for $\beta=1/4$ displaying the correct asymptotic behavior for large $k$ is also known [13],

$$H_{1/4}(k) = \frac{\tau_0}{8\pi(k\tau_0)^{5/4}} \int_0^\infty u^{-3/4} \exp\left[-\frac{1}{4}\left(\frac{1}{\sqrt{k\tau_0 u}} + u\right)\right] du. \tag{15}$$

The probability density function associated to the Kohlrausch decay function is plotted in Figure 1 for selected values of the parameter $\beta$.

**Figure 1**

As mentioned, the stretched exponential decay function has an undesirable short-time behavior (infinite initial rate, faster-than-exponential decay for short times). For this reason, a modified form was proposed [11]

$$I(t) = \exp\left[\alpha^\beta - \left(\alpha + \frac{t}{\tau_0}\right)^\beta\right], \tag{16}$$

where $\alpha$ is a nonnegative dimensionless parameter. The time-dependent rate coefficient is in this case

$$w(t) = \frac{\dfrac{\beta}{\tau_0}}{\left(\alpha + \dfrac{t}{\tau_0}\right)^{1-\beta}} \quad . \tag{17}$$

## 3. Compressed hyperbola (or Becquerel) function

The first quantitative studies of the time evolution of luminescence (following flash excitation), were carried out by Edmond Becquerel (1820-1891) and published in 1861. The functions used by this author for the description of the experimental decays already included an exponential of time, and also a sum of two such exponentials [14]. Becquerel also noticed that for some of his experimental systems (with inorganic solids), an empirical decay function of the form

$$I(t) = \frac{1}{(1+at)^2} \quad , \tag{18}$$

gave better fits than a sum of two exponentials. Later on, he proposed a more general equation in the form $I^m(t+a) = b$ that can be rewritten as

$$I(t) = \frac{1}{(1+at)^p} \quad , \tag{19}$$

with $p$ taking values between 1 and 2 [14].

This function that decays faster than a hyperbola (for which $p=1$) can be called *compressed hyperbola*. Owing to Becquerel pioneering studies, this function is also called the Becquerel decay function [15]. The Becquerel decay function can be rewritten as [15],

$$I(t) = \frac{1}{\left[1 + \alpha \dfrac{t}{\tau_0}\right]^{\frac{1}{\alpha}}}, \tag{20}$$

where $0 \leq \alpha \leq 1$. The restriction p≤2 is thus lifted. Values of $\alpha$ outside the range defined yield unphysical results: for $\alpha \geq 1$ the integrated intensity (total intensity) diverges, and for $\alpha < 0$ the intensity becomes zero at a finite value of *t*.

The corresponding time-dependent rate coefficient is [15]

$$w(t) = \frac{\dfrac{1}{\tau_0}}{1 + \alpha \dfrac{t}{\tau_0}}. \qquad (21)$$

The Becquerel decay function possesses a simple inverse Laplace transform [15],

$$H(k) = \frac{\tau_0}{\alpha \, \Gamma\!\left(\dfrac{1}{\alpha}\right)} \left(\frac{k\tau_0}{\alpha}\right)^{\frac{1}{\alpha}-1} \exp\!\left(-\frac{k\tau_0}{\alpha}\right), \qquad (22)$$

which is a Gamma distribution, and whose mean and standard deviation are $1/\tau_0$ and $\sqrt{\alpha}/\tau_0$, respectively. This probability density function is plotted in Figure 2 for selected values of the parameter $\alpha$. For $\alpha=1$, the distribution of rate constants is exponential. On the other hand, for $\alpha$ sufficiently close to 0, the rate constant distribution becomes a relatively narrow normal (Gaussian) distribution with mean and standard deviation given by the previous expressions. When $\alpha \to 0$, the standard deviation goes to zero, and the distribution becomes $\delta\!\left(k - \dfrac{1}{\tau_0}\right)$.

**Figure 2**

The Becquerel decay is initially exponential,

$$I(t) = \exp\!\left(-\frac{t}{\tau_0}\right), \qquad (23)$$

and the initial decay rate is always finite. This short-time behavior is more realistic than that displayed by the primitive Kohlrausch function, but is shared by the modified Kohlrausch function mentioned.

The Becquerel function is a quite flexible decay function, although its less direct relation to the exponential decay has limited its use up to now mainly to the luminescence of phosphors [15]. Nevertheless, there are some applications in fluorescence. For instance, Wlodarczyk and Kierdaszuk [16] showed that it provides good fits for fluorescence decays that slightly depart from the exponential behavior, implying a relatively narrow distribution of decay times around a mean value.

## 4. A more general decay function

Comparison of equations (17) and (21) suggests a time-dependent rate coefficient encompassing both the stretched exponential and the compressed hyperbola,

$$w(t) = \frac{\frac{1}{\tau_0}}{\left(1 + \alpha \frac{t}{\tau_0}\right)^{1-\beta}}, \tag{24}$$

where α and β are dimensionless parameters. Indeed, with this rate coefficient equation (2) yields

$$I(t) = \exp\left[\frac{1 - \left(1 + \alpha \frac{t}{\tau_0}\right)^{\beta}}{\alpha\beta}\right], \tag{25}$$

which for short times reduces to an exponential function with lifetime $\tau_0$, cf. equation (24). For very small α, the decay is also close to single exponential for all times, and irrespective of β.

On the other hand, this relaxation function gives, for sufficiently long times (how long they need to be depends on α) the stretched exponential form,

$$I(t) = \exp\left[-\left(\frac{t}{\tau_0'}\right)^{\beta}\right], \tag{26}$$

with

$$\tau_0' = \frac{(\alpha\beta)^{1/\beta}}{\alpha} \tau_0, \tag{27}$$

except for $\beta = 0$, where equation (25) becomes for all times

$$I(t) = \frac{1}{\left(1 + \alpha \frac{t}{\tau_0}\right)^{1/\alpha}}, \tag{28}$$

which is the compressed hyperbola function if 0<α<1, and the stretched hyperbola function if α>1 (in the last case the integrated decay diverges).

The function given by equation (25) encompasses thus the stretched exponential and the compressed hyperbola relaxation functions in a much simpler manner than that of a previous proposal [15].

The distribution of rate constants corresponding to equation (25) can be obtained by means of equations (5)-(7), with

$$\text{Re}[I(i\omega)] = \exp\left[\frac{1-\rho^\beta \cos(\beta\theta)}{\alpha\beta}\right]\cos\left(\frac{\rho^\beta \sin(\beta\theta)}{\alpha\beta}\right), \tag{29}$$

$$\text{Im}[I(i\omega)] = -\exp\left[\frac{1-\rho^\beta \cos(\beta\theta)}{\alpha\beta}\right]\sin\left(\frac{\rho^\beta \sin(\beta\theta)}{\alpha\beta}\right), \tag{30}$$

where

$$\rho = \sqrt{1+\left(\frac{\alpha\omega}{\tau_0}\right)^2}, \tag{31}$$

and

$$\theta = \arctan\left(\frac{\alpha\omega}{\tau_0}\right). \tag{32}$$

The probability density function of rate constants, $H_{\alpha\beta}(k)$, obtained by numerical integration, is shown in Figures 3 and 4 for selected values of the parameters.

**Figure 3**

In Figure 3, parameter $\alpha$ is fixed at 1. It is seen that the probability density function $H_{1\beta}(k)$ already closely follows the trend with $\beta$ displayed by the stretched exponential one, compare Figure 1.

**Figure 4**

In Figure 4, parameter $\alpha$ is fixed at 0.01. It is seen that the probability density function $H_{0.01\beta}(k)$ is always narrow and peaks at *k=1*, irrespective of $\beta$, being nearly Gaussian for $\beta$ close to 0 and approaching $\delta(k\text{-}1)$ as $\beta\rightarrow 1$.

## 5. Conclusions

A relaxation function $I(t/\tau_0; \alpha, \beta)$ unifying the stretched exponential with the compressed hyperbola was obtained, equation (25), and its properties studied. The scaling parameter $\tau_0$ has dimensions of time, whereas the shape-determining parameters $\alpha$ and $\beta$ are dimensionless, with $1>\alpha>0$ and $1>\beta>0$. For short times, the relaxation function is always exponential, with time $\tau_0$. For small values of $\alpha$, the function is close to exponential for all times, irrespective of $\beta$. The function is also close to an exponential when $\beta$ is near unity, irrespective of $\alpha$. For large values of $\alpha$ and long times, the function is close to a stretched exponential, provided $\beta>0$. The compressed hyperbola is recovered for $\beta=0$.

The interest of function $I(t/\tau_0; \alpha, \beta)$ results from three aspects: Firstly, it has a simple mathematical form, appropriate for data fitting; Secondly, it demonstrates the continuity between two apparently disparate relaxation functions, the stretched exponential and the compressed hyperbola; Thirdly, and finally, it can in principle be used to describe a class of complex relaxations processes that do not follow exactly neither the stretched nor the hyperbolic laws.

**Figure captions**

**Fig. 1** Distribution of rate constants (probability density function) for the stretched exponential (or Kohlrausch) relaxation function. The number next to each curve is the respective $\beta$.

**Fig. 2** Distribution of rate constants (probability density function) for the compress hyperbola (or Becquerel) relaxation function. The number next to each curve is the respective $\alpha$.

**Fig. 3** Distribution of rate constants (probability density function) for the general relaxation function, equation (25), as a function of both $k$ and parameter $\beta$. The parameter $\alpha$ is fixed at 1.

**Fig. 4** Distribution of rate constants (probability density function) for the general relaxation function, equation (25), as a function of both $k$ and parameter $\beta$. The parameter $\alpha$ is fixed at 0.01.

**Fig. 1**

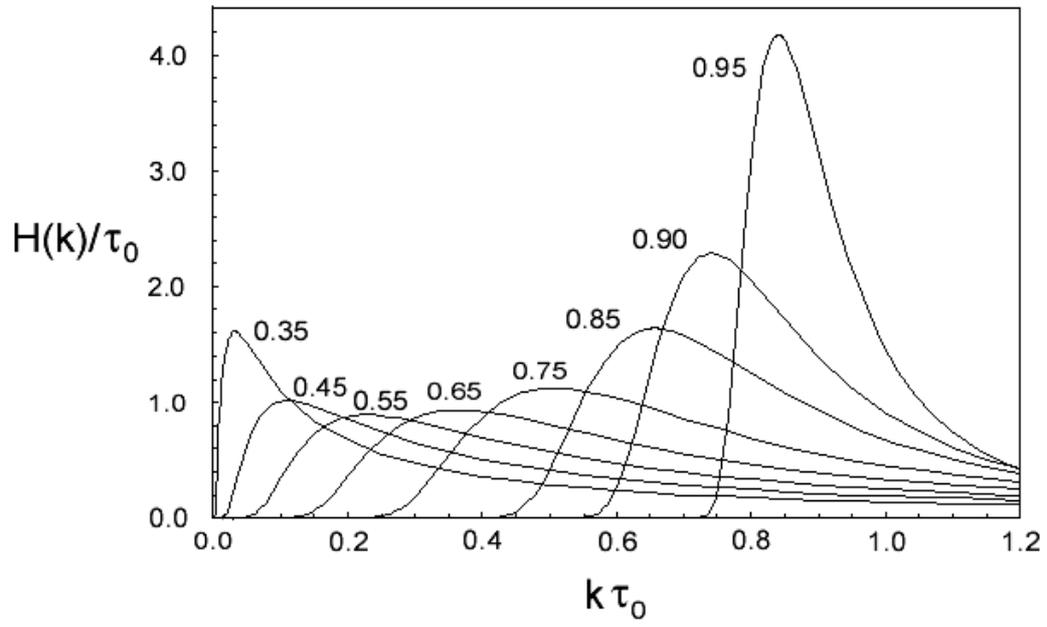

**Fig. 2**

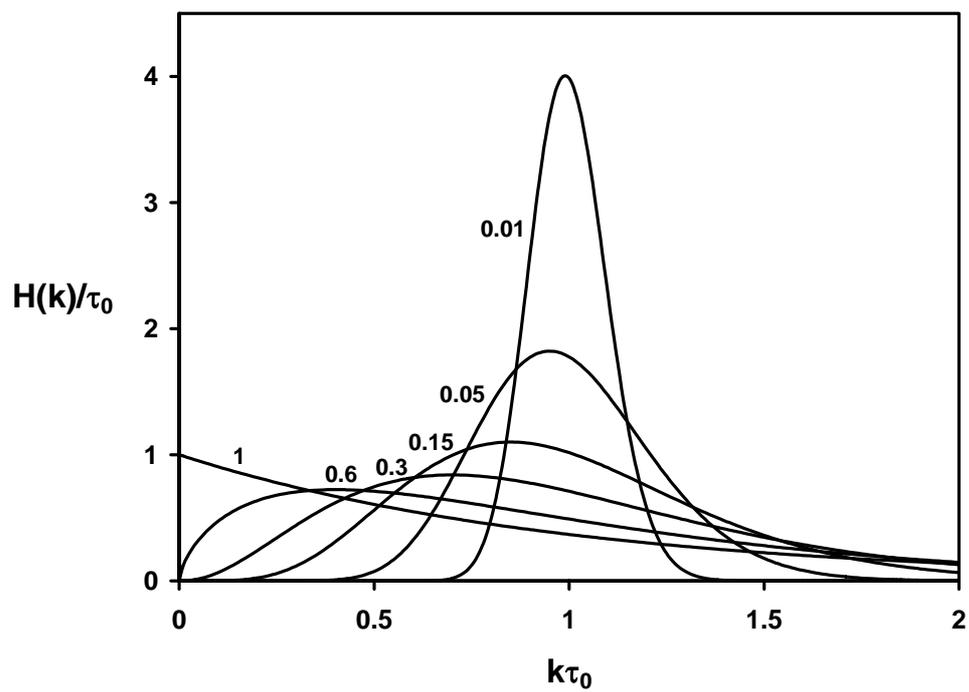

**Fig. 3**

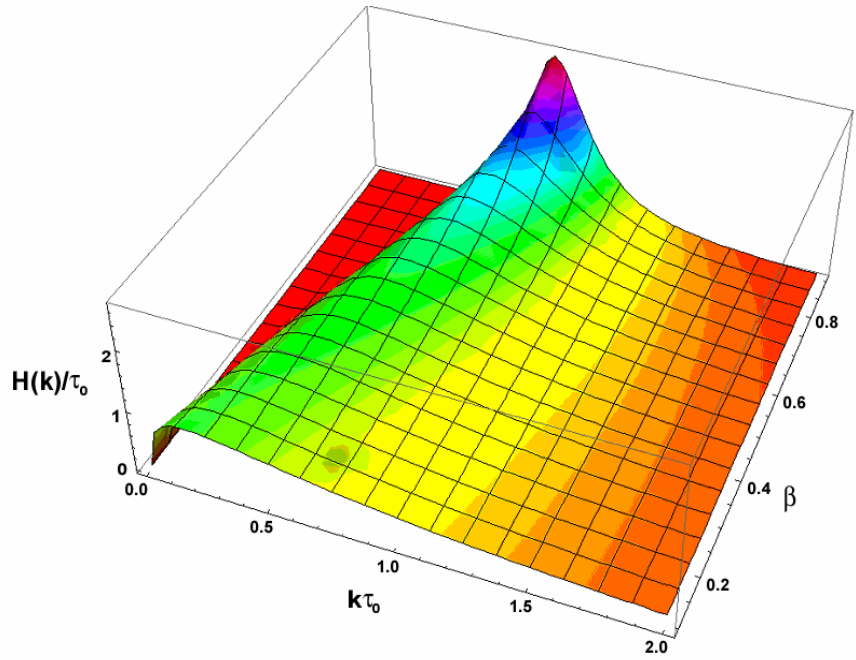

**Fig. 4**

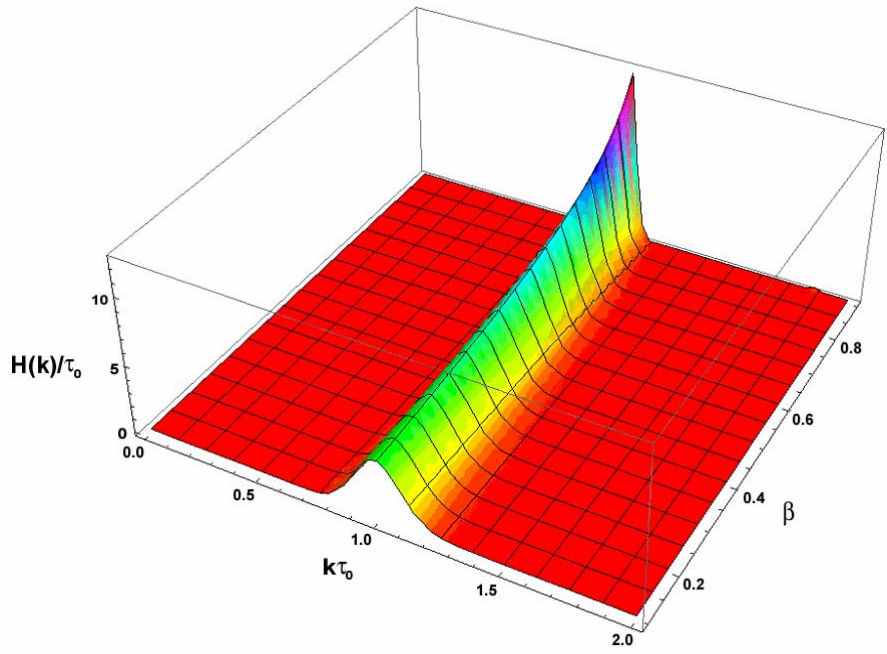